\documentclass[12pt]{article}
\usepackage{inputenc,mathtools,booktabs,indentfirst,authblk,geometry,graphicx}

\title{\textbf{\textit{The application of adaptive perturbation theory in strongly coupled double harmonic oscillator system}}}
\author{Xin Guo\footnote{email address: guoxinam@outlook.com or guoxin@m.scnu.edu.cn}}
\affil{Institute of Quantum Matter,
	School of Physics and Telecommunication Engineering,
	South China Normal University, Guangzhou 510006, Guangdong, China.}

\date{Sep 2020} 

\begin{document}
	\maketitle
	
	\tableofcontents
	\newpage
	
	\begin{center}
		{\Large \textbf{Abstract}} 
	\end{center}
	
	The idea of adaptive perturbation theory is to divide a Hamiltonian into a solvable part and a perturbation part. The solvable part contains the non-interacting sector and  the diagonal elements of Fock space from the interacting terms. The perturbed term is the non-diagonal sector of Fock space. Therefore, the perturbation parameter is not coupling constant. This is different from the standard procedure of previous perturbation method. In this letter, we use the adaptive perturbation theory to extract the solvable elements in the strongly coupled double harmonic oscillator system and obtain the energy spectrum of the solvable part. Then, we diagonalize the Hamiltonian to obtain the numerical solution.  In order to study the accuracy of adaptive perturbation theory in the strongly coupled double harmonic oscillator system, we demonstrate the analytical study from the leading order and second-order perturbation. The deviations between the leading-order and the numerical solution show that as the gap between the quasiparticle number \textbf{\textit{$n_{1}$}} and \textbf{\textit{$n_{2}$}} decreases, or the coupling constant \textbf{\textit{$\lambda$}} increases, the gap between the exact solutions in the strongly coupling region of double harmonic oscillator system and the numerical solutions becomes smaller. Overall, the numerical gap is about 1\% to 3\%, which is not a bad result. The value of the second-order is quite close to the numerical solution when \textbf{\textit{$n_{1}$}} doesn't equal \textbf{\textit{$n_{2}$}}.  In most cases, the deviation is less than 1\%, which means that the adaptive perturbation theory is effective for the double harmonic oscillator system in strong coupling field. \\
	
	
	\section{Introduction}
	{\indent
		Until now, people have only obtained the exact results of several quantum models, such as hydrogen atom and quantum harmonic oscillator. Unfortunately, the potential fields of these quantum models are too idealized to describe most of the more complex quantum systems. But a systematic theory, called perturbation theory, can solve more complex quantum systems based on the exact solutions of these ideal quantum models. Therefore, perturbation theory is an important theory of quantum mechanics and has become the focus of many physicists.}\\
	
	The procedure of perturbation method is to begin from the non-interacting system and then do a perturbation from the coupling terms. Therefore, people cannot apply the procedure to the strongly coupling region. As for Quantum Field Theory(QFT), we encounter challenges in both strong coupling and weak coupling regions. When it comes to strong coupling, we have to mention Quantum Chromodynamics (QCD), which is a subject to describe the interaction between quarks and gluons. To study the strongly coupled QCD, we need to study new perturbation theory for developing quantum field theory in strong coupling.  After studying the Higgs particle, we came to a conclusion that the perturbation method is a useful tool for studying the fundamental physics[1]. But the eigenstate of the vanishing coupling constant case in the perturbation method does not meet the requirement that it is should be a Fock state[1], so we have to find a new method for the problems related to Higgs particle. In addition, due to the strongly coupled QFT only can be studied by the lattice method, it is hard to know whether the lattice method is correct due to the continuum limit[1]. We need another way to verify. Finally, although the lattice method is very good, the current computer resources can not meet its requirements. \\
	
	As an unique perturbation method, the adaptive perturbation theory is proposed to solve the above problems. The idea of the new theory is to divide a Hamiltonian into a solvable part and a perturbation part. How to divide these two parts is the key to its difference from the previous perturbation theory and the essence of this idea [2]. The solvable part consists of two parts: the non-interacting sector and the diagonal elements of the Fock space from the interacting terms. The perturbed term is the non-diagonal sector of the Fock space. Therefore, the perturbation parameter is not coupling constant[1]. In the following sections, we study the accuracy of adaptive perturbation theory in the strongly coupled double harmonic oscillator system. We demonstrate an analytical study from the leading-order and second-order perturbation. The corresponding results are compared with the numerical solutions. At the leading order, in order to define the quasiparticle number \textbf{\textit{$n$}} of the strongly coupled double harmonic oscillator system in the adaptive perturbation theory, we introduce a variable $\gamma$. This variable is carefully selected according to the requirement of not breaking the commutation relation between momentum and position operators, and its value is determined by minimizing the expectation value of the energy. At the second order, we also choose the $\gamma$ of the leading-order to calculate the energy. The final result should not depend on the choice of the $\gamma$ if we do the perturbation to all orders[3]. Finally, we analyze the reason why the adaptive perturbation theory can converge, and also look forward to the research areas when the adaptive perturbation theory is extended to the strong coupling quantum field theory. \\
	
	\textbf{\textit{In this letter, we want to verify whether the adaptive perturbation theory can be applied to the strong coupling double harmonic oscillator quantum system, and how the effect is. The reasons for studying this problem are as follows:}} \\
	
	\textbf{\textit{We know that when quantum field theory is in the field of strong coupling, it is usually studied by lattice method. Due to the limitation of continuum, the correctness of lattice method cannot be guaranteed. In some cases, even when the coupling constant is very small, the perturbation term is only asymptotically convergent. Therefore, it is imperative to develop new perturbation theory. Since quantum field theory contains innumerable coupled harmonic oscillators, if we want to study the perturbation theory applicable to the strong coupling quantum field theory, we should first study the application of this theory in the strong coupling double harmonic oscillator system. In order to develop the perturbation theory which can be applied to the strong coupling quantum field theory, we want to verify whether the adaptive perturbation theory can be applied to the strong coupling double harmonic oscillator quantum system, and how the effect is. That's what we want to do in this article.}}
	
	\section{Adaptive Perturbation Method}
	\subsection{Leading-Order}
	
	The strongly coupled double harmonic oscillator system is defined by a Hamiltonian of the form($\lambda$ equals $\lambda^{\prime}$. The derivation of formulas in this chapter does not consider dimensions)
	\begin{align}
	H=\frac{1}{2}p_1^2+\frac{1}{2}p_2^2+\frac{1}{6}\lambda(x_1^4+x_2^4)+\frac{1}{3}\lambda^{\prime}(x_1^2+x_2^2)+\lambda x_1^2x_2^2
	\end{align}
	
	\noindent where \textbf{\textit{$\lambda$}} is the coupling constant, \textbf{\textit{p}} is the canonical momentum, \textbf{\textit{x}} is the canonical position satisfying the commutation relation
	
	\begin{align}
	[x,p]=i
	\end{align}
	
	We use formula 4 and formula 5 to define $A_{\gamma}$, $A^{\dagger}_{\gamma}$ for decomposing a
	Hamiltonian into a solvable part and a perturbation part, and they satisfy the following relationship [2]
	
	\begin{align}
	[A_{\gamma},A^{\dagger}_{\gamma}]=1
	\end{align}
	
	\begin{align}
	x=\frac{1}{\sqrt{2\gamma}}(A^{\dagger}_{\gamma}+A_{\gamma})
	\end{align}
	
	\begin{align}
	p=i\sqrt{\frac{\gamma}{2}}(A^{\dagger}_{\gamma}-A_{\gamma})
	\end{align}
	
	Define \textbf{\textit{$\gamma$}}-dependent decomposition of \textbf{\textit{H}} into an unperturbed part and a perturbation as follows
	
	\begin{align}
	H=&\;H_{0}(\gamma_1\gamma_2)+V(\gamma_1\gamma_2) 
	\\
	H_{0}(\gamma_1\gamma_2)=&\;(\frac{\gamma_1}{4}+\frac{\lambda}{4\gamma_1^2}+\frac{\lambda}{6\gamma_1})(2A_{\gamma_1}^{\dagger}A_{\gamma_1}+\frac{1}{2}) 
	\\
	&+\frac{\lambda}{4\gamma_1^2}A_{\gamma_1}^{\dagger}A_{\gamma_1}(A_{\gamma_1}^{\dagger}A_{\gamma_1}-1)+\frac{\gamma_1}{8}+\frac{\lambda}{12\gamma_1} \notag
	\\
	&+(\frac{\gamma_2}{4}+\frac{\lambda}{4\gamma_2^2}+\frac{\lambda}{6\gamma_2})(2A_{\gamma_2}^{\dagger}A_{\gamma_2}+\frac{1}{2}) \notag
	\\
	&+\frac{\lambda}{4\gamma_2^2}A_{\gamma_2}^{\dagger}A_{\gamma_2}(A_{\gamma_2}^{\dagger}A_{\gamma_2}-1)+\frac{\gamma_2}{8}+\frac{\lambda}{12\gamma_2} \notag
	\\
	&+\frac{\lambda}{4\gamma_1\gamma_2}(4A_{\gamma_1}^{\dagger}A_{\gamma_1}A_{\gamma_2}^{\dagger}A_{\gamma_2}+2A_{\gamma_1}^{\dagger}A_{\gamma_1}+2A_{\gamma_2}^{\dagger}A_{\gamma_2}+1) \notag
	\\
	V(\gamma_1\gamma_2)=&\;(-\frac{\gamma_1}{4}+\frac{\lambda}{4\gamma_1^2}+\frac{\lambda}{6\gamma_1})(A_{\gamma_1}^{\dagger2}+A_{\gamma_1}^{2}) 
	\\
	&+\frac{\lambda}{4\gamma_1^2}\{\frac{1}{6}(A_{\gamma_1}^{\dagger4}+A_{\gamma_1}^{4})+\dfrac{2}{3}(A_{\gamma_1}^{\dagger3}A_{\gamma_1}+A_{\gamma_1}^{\dagger}A_{\gamma_1}^{3})\} \notag
	\\
	&+(-\frac{\gamma_2}{4}+\frac{\lambda}{4\gamma_2^2}+\frac{\lambda}{6\gamma_2})(A_{\gamma_2}^{\dagger2}+A_{\gamma_2}^{2}) \notag
	\\
	&+\frac{\lambda}{4\gamma_2^2}\{\frac{1}{6}(A_{\gamma_2}^{\dagger4}+A_{\gamma_2}^{4})+\dfrac{2}{3}(A_{\gamma_2}^{\dagger3}A_{\gamma_2}+A_{\gamma_2}^{\dagger}A_{\gamma_2}^{3})\} \notag
	\\
	&+\frac{\lambda}{4\gamma_1\gamma_2}(A_{\gamma_1}^{\dagger2}A_{\gamma_2}^{\dagger2}+A_{\gamma_1}^{\dagger2}A_{\gamma_2}^{2}+2A_{\gamma_1}^{\dagger2}A_{\gamma_2}^{\dagger}A_{\gamma_2} \notag
	\\
	&+A_{\gamma_1}^{\dagger2}+A_{\gamma_1}^{2}A_{\gamma_2}^{\dagger2}+A_{\gamma_1}^{2}A_{\gamma_2}^{2}+2A_{\gamma_1}^{2}A_{\gamma_2}^{\dagger}A_{\gamma_2}+A_{\gamma_1}^{2} \notag
	\\
	&+2A_{\gamma_1}^{\dagger}A_{\gamma_1}A_{\gamma_2}^{\dagger2}+2A_{\gamma_1}^{\dagger}A_{\gamma_1}A_{\gamma_2}^{2}+A_{\gamma_2}^{\dagger2}+A_{\gamma_2}^{2}) \notag
	\end{align}
	
	$|0_{\gamma}\rangle$ is the \textbf{\textit{$\gamma$}}-dependent vacuum state [2], which means the 0-particle state. $|n_{\gamma}\rangle$ is the n-particle state built on this vacuum state.
	
	\begin{align}
	A_{\gamma}|0_{\gamma}\rangle=0
	\end{align}
	
	So
	
	\begin{align}
	A^{\dagger}_{\gamma}A_{\gamma}\frac{1}{\sqrt{n!}}A^{\dagger n}_{\gamma}|0_{\gamma}\rangle
	=n\frac{1}{\sqrt{n!}}A^{\dagger n}_{\gamma}|0_{\gamma}\rangle
	\end{align}
	
	Define 
	
	\begin{align}
	N_{\gamma}=A^{\dagger}_{\gamma}A_{\gamma} 
	\end{align}
	
	\begin{align}
	|n_{\gamma}\rangle=\frac{1}{\sqrt{n!}}A^{\dagger n}_{\gamma}|0_{\gamma}\rangle
	\end{align}
	
	\begin{align}
	N_{\gamma}|n_{\gamma}\rangle=n|n_{\gamma}\rangle
	\end{align}
	
	The value of \textbf{\textit{$\gamma$}}(\textgreater 0) used to define the quasiparticle number \textbf{\textit{$n$}} of the strongly coupled double harmonic oscillator system is determined by requiring that it minimizes the expectation value of the leading-order[2]
	
	\begin{align}
	E_{n_1,n_2}(\gamma_1\gamma_2)=&<n_{\gamma_1\gamma_2}|H|n_{\gamma_1\gamma_2}> 
	\\
	=&<N_{\gamma_1\gamma_2}|H_{0}(\gamma_1\gamma_2)|n_{\gamma_1\gamma_2}> \notag
	\\
	=&(\frac{\gamma_1}{4}+\frac{\lambda}{4\gamma_1^2}+\frac{\lambda}{6\gamma_1})(2n_1+\frac{1}{2}) \notag
	\\
	&+\frac{\lambda}{4\gamma_1^2}n_1(n_1-1)+\frac{\gamma_1}{8}+\frac{\lambda}{12\gamma_1} \notag
	\\
	&+(\frac{\gamma_2}{4}+\frac{\lambda}{4\gamma_2^2}+\frac{\lambda}{6\gamma_2})(2n_2+\frac{1}{2}) \notag
	\\
	&+\frac{\lambda}{4\gamma_2^2}n_2(n_2-1)+\frac{\gamma_2}{8}+\frac{\lambda}{12\gamma_2} \notag
	\\
	&+\frac{\lambda}{4\gamma_1\gamma_2}(2n_1+1)(2n_2+1) \notag
	\end{align}
	
	Minimizing $E(\gamma_1\gamma_2)$ with respect to $\gamma$ leads to the equation
	\begin{align}
	\frac{\partial E_{n_1,n_2}(\gamma_1\gamma_2)}{\partial \gamma_1}=&\frac{\partial }{\partial \gamma_1}\{(\frac{\gamma_1}{4}+\frac{\lambda}{4\gamma_1^2}+\frac{\lambda}{6\gamma_1})(2n_1+\frac{1}{2}) 
	\\
	&+\frac{\lambda}{4\gamma_1^2}n_1(n_1-1)+\frac{\gamma_1}{8}+\frac{\lambda}{12\gamma_1} \notag
	\\
	&+(\frac{\gamma_2}{4}+\frac{\lambda}{4\gamma_2^2}+\frac{\lambda}{6\gamma_2})(2n_2+\frac{1}{2}) \notag
	\\
	&+\frac{\lambda}{4\gamma_2^2}n_2(n_2-1)+\frac{\gamma_2}{8}+\frac{\lambda}{12\gamma_2} \notag
	\\
	&+\frac{\lambda}{4\gamma_1\gamma_2}(2n_1+1)(2n_2+1)\} \notag
	\\
	=&(\frac{1}{4}-\frac{\lambda}{2\gamma_1^3}-\frac{\lambda}{6\gamma_1^2})(2n_1+\frac{1}{2}) \notag
	\\
	&-\frac{\lambda}{2\gamma_1^3}n_1(n_1-1)+\frac{1}{8}-\frac{\lambda}{12\gamma_1^2} \notag
	\\
	&-\frac{\lambda}{4\gamma_1^2\gamma_2}(2n_1+1)(2n_2+1) \notag
	\\
	=&0 \notag
	\end{align}
	
	\begin{align}
	\frac{\partial E_{n_1,n_2}(\gamma_1\gamma_2)}{\partial\gamma_2}=&(\frac{1}{4}-\frac{\lambda}{2\gamma_2^3}-\frac{\lambda}{6\gamma_2^2})(2n_2+\frac{1}{2}) 
	\\
	&-\frac{\lambda}{2\gamma_2^3}n_2(n_2-1)+\frac{1}{8}-\frac{\lambda}{12\gamma_2^2} \notag
	\\
	&-\frac{\lambda}{4\gamma_1\gamma_2^2}(2n_1+1)(2n_2+1) \notag
	\\
	=&0 \notag
	\end{align}
	
	In order to obtain the numerical solutions for comparison with the results of adaptive perturbation theory, we use the naive discretization for the kinematic term [4]
	
	\begin{align}
	p^2\psi\longrightarrow -\frac{\psi_{j+1}-2\psi_{j}+\psi_{j-1}}{a^2}
	\end{align}
	
	\noindent where \textbf{\textit{a}} is the lattice spacing, and \textbf{\textit{$\psi_{j}$}} is the eigenfunction for the lattice. The lattice index is labeled by\textbf{\textit{ j}} = 1, 2, · · · , \textbf{\textit{n}}, where \textbf{\textit{n}} = 512 is the number of lattice points. The lattice size is marked with \textbf{\textit{L}}
	
	\begin{align}
	L=8\equiv \frac{na}{2}
	\end{align}
	
	To obtain the numerical solution, the exact diagolization is done.

    \subsection{2nd-Order}	
   	The adaptive perturbation theory is also suitable for the energy correction formula of the time-independent perturbation
    
    \begin{align}
    E_n=E_n^{(0)}+\langle n^{(0)}|V|n^{(0)}\rangle+\sum_{k \neq n}\dfrac{|\langle k^{(0)}|V|n^{(0)}\rangle|^2}{E_n^{(0)}-E_k^{(0)}}+...
    \end{align}
    
    \noindent where $E_n^{(0)}$ is the n-th unperturbed eigenenergy, $|n^{(0)}\rangle$ is the n-th unperturbed eigenstate, and $E_k^{(0)}$ is the k-th unperturbed eigenenergy, calculated by the n-th unperturbed eigenstate’s $\gamma$[1].
    
    It is applied to the strongly coupled double harmonic oscillator system, and the formula is as follows
    
    \begin{align}
    E_{n_1,n_2}=&E_{n_1,n_2}^{(0)}+\langle {(n_1,n_2)}^{(0)}|V|{(n_1,n_2)}^{(0)}\rangle \\
    &+\sum_{(k_1,k_2)\neq(n_1,n_2)}\dfrac{|\langle (k_1,k_2)^{(0)}|V|(n_1,n_2)^{(0)}\rangle|^2}{E_{n_1,n_2}^{(0)}-E_{k_1,k_2}^{(0)}}+... \notag
    \end{align}
    
    The first-order term $\langle {(n_1,n_2)}^{(0)}|V|{(n_1,n_2)}^{(0)}\rangle$ vanishes due to that V is a
    non-diagonal element of the Fock space[1]. 
    
    \hspace*{\fill}
    
    For the convenience of writing, we introduce the following parameters
    
    \begin{align}
    A_1&=-\frac{\gamma_1}{4}+\frac{\lambda}{4\gamma_1^2}+\frac{\lambda}{6\gamma_1} 
    \\
    A_2&=\frac{\lambda}{24\gamma_1^2} \notag
    \\
    A_3&=\frac{\lambda}{6\gamma_1^2}  \notag
    \\
    B_1&=-\frac{\gamma_2}{4}+\frac{\lambda}{4\gamma_2^2}+\frac{\lambda}{6\gamma_2} \notag
    \\
    B_2&=\frac{\lambda}{24\gamma_2^2} \notag
    \\
    B_3&=\frac{\lambda}{6\gamma_2^2} \notag
    \\
    C&=\frac{\lambda}{4\gamma_1\gamma_2} \notag
    \end{align}
    
    The necessary of the transition energy is given as in the following 
    
    \begin{align}
    &E_{n_1,n_2}^{(0)}(\gamma_1\gamma_2)-E_{n_1-2,n_2-2}^{(0)}(\gamma_1\gamma_2)
    \\
    =&\frac{1}{6\gamma_1^2\gamma_2^2}(3\lambda(-1+2n_1)\gamma_2^2+6\gamma_1^3\gamma_2^2+4\lambda \gamma_1\gamma_2(-3+3n_1+3n_2+\gamma_2) \notag
    \\
    &+\gamma_1^2(-3\lambda+6\lambda n_2+4\lambda \gamma_2+6\gamma_2^3)) \notag
    \end{align}
    
    \begin{align}
    &E_{n_1,n_2}^{(0)}(\gamma_1\gamma_2)-E_{n_1-2,n_2+2}^{(0)}(\gamma_1\gamma_2)
    \\
    =&\frac{1}{6\gamma_1^2\gamma_2^2}(3\lambda(-1+2n_1)\gamma_2^2+6\gamma_1^3\gamma_2^2+4\lambda \gamma_1\gamma_2(6-3n_1+3n_2+\gamma_2) \notag
    \\
    &-\gamma_1^2(9\lambda+6\lambda n_2+4\lambda \gamma_2+6\gamma_2^3)) \notag
    \end{align}

	\begin{align}
	&E_{n_1,n_2}^{(0)}(\gamma_1\gamma_2)-E_{n_1-2,n_2}^{(0)}(\gamma_1\gamma_2)
	\\
	=&\frac{1}{6\gamma_1^2\gamma_2}(3\lambda(-1+2n_1)\gamma_2+6\gamma_1^3\gamma_2+2\lambda \gamma_1(3+6n_2+2\gamma_2)) \notag
	\end{align}
	
    \begin{align}
	&E_{n_1,n_2}^{(0)}(\gamma_1\gamma_2)-E_{n_1+2,n_2-2}^{(0)}(\gamma_1\gamma_2)
	\\
	=&\frac{1}{6\gamma_1^2\gamma_2^2}(-3\lambda(3+2n_1)\gamma_2^2-6\gamma_1^3\gamma_2^2-4\lambda \gamma_1\gamma_2(-6-3n_1+3n_2+\gamma_2) \notag
	\\
	&+\gamma_1^2(-3\lambda+6\lambda n_2+4\lambda \gamma_2+6\gamma_2^3)) \notag
	\end{align}
	   
    \begin{align}
    &E_{n_1,n_2}^{(0)}(\gamma_1\gamma_2)-E_{n_1+2,n_2+2}^{(0)}(\gamma_1\gamma_2)
    \\
    =&-\frac{1}{6\gamma_1^2\gamma_2^2}(3\lambda(3+2n_1)\gamma_2^2+6\gamma_1^3\gamma_2^2+4\lambda \gamma_1\gamma_2(9+3n_1+3n_2+\gamma_2) \notag
    \\
    &+\gamma_1^2(9\lambda+6\lambda n_2+4\lambda \gamma_2+6\gamma_2^3)) \notag
    \end{align}
   
	\begin{align}
	&E_{n_1,n_2}^{(0)}(\gamma_1\gamma_2)-E_{n_1+2,n_2}^{(0)}(\gamma_1\gamma_2)
	\\
	=&-\frac{1}{6\gamma_1^2\gamma_2}(3\lambda(3+2n_1)\gamma_2+6\gamma_1^3\gamma_2+2\lambda \gamma_1(3+6n_2+2\gamma_2)) \notag
	\end{align}   
   
	\begin{align}
	&E_{n_1,n_2}^{(0)}(\gamma_1\gamma_2)-E_{n_1,n_2-2}^{(0)}(\gamma_1\gamma_2)
	\\
	=&\frac{1}{6\gamma_1\gamma_2^2}(6\lambda(1+2n_1)\gamma_2+\gamma_1(-3\lambda+6\lambda n_2+4\lambda\gamma_2+6\gamma_2^3)) \notag
	\end{align}   
   
	\begin{align}
	&E_{n_1,n_2}^{(0)}(\gamma_1\gamma_2)-E_{n_1,n_2+2}^{(0)}(\gamma_1\gamma_2)
	\\
	=&-\frac{1}{6\gamma_1\gamma_2^2}(6\lambda(1+2n_1)\gamma_2+\gamma_1(9\lambda+6\lambda n_2+4\lambda\gamma_2+6\gamma_2^3)) \notag
	\end{align}     
   
	\begin{align}
	&E_{n_1,n_2}^{(0)}(\gamma_1\gamma_2)-E_{n_1-4,n_2}^{(0)}(\gamma_1\gamma_2)
	\\
	=&\frac{1}{3\gamma_1^2\gamma_2}(3\lambda(-3+2n_1)\gamma_2+6\gamma_1^3\gamma_2+2\lambda\gamma_1(3+6n_2+2\gamma_2)) \notag
	\end{align}   
  
	\begin{align}
	&E_{n_1,n_2}^{(0)}(\gamma_1\gamma_2)-E_{n_1+4,n_2}^{(0)}(\gamma_1\gamma_2)
	\\
	=&-\frac{1}{3\gamma_1^2\gamma_2}(3\lambda(5+2n_1)\gamma_2+6\gamma_1^3\gamma_2+2\lambda\gamma_1(3+6n_2+2\gamma_2)) \notag
	\end{align}     
  
	\begin{align}
	&E_{n_1,n_2}^{(0)}(\gamma_1\gamma_2)-E_{n_1,n_2-4}^{(0)}(\gamma_1\gamma_2)
	\\
	=&\frac{1}{3\gamma_1\gamma_2^2}(6\lambda(1+2n_1)\gamma_2+\gamma_1(-9\lambda+6\lambda n_2+4\lambda\gamma_2+6\gamma_2^3)) \notag
	\end{align}  
	  
	\begin{align}
	&E_{n_1,n_2}^{(0)}(\gamma_1\gamma_2)-E_{n_1,n_2+4}^{(0)}(\gamma_1\gamma_2)
	\\
	=&-\frac{1}{3\gamma_1\gamma_2^2}(6\lambda(1+2n_1)\gamma_2+\gamma_1(15\lambda+6\lambda n_2+4\lambda\gamma_2+6\gamma_2^3)) \notag
	\end{align}  

    $E_{n_1,n_2}^{(0)}(\gamma_1\gamma_2)$ is defined by $E_{n_1,n_2}(\gamma_1\gamma_2)_{min}$, and the second-order perturbation gives
    
    \begin{align}
    E_{n_1,n_2}(\gamma_1\gamma_2)_2=E_{n_1,n_2}^{(0)}&(\gamma_1\gamma_2) \\
    +&\sum_{(k_1,k_2)\neq(n_1,n_2)}\dfrac{|\langle (k_1,k_2)^{(0)}|V|(n_1,n_2)^{(0)}\rangle|^2}{E_{n_1,n_2}^{(0)}(\gamma_1\gamma_2)-E_{k_1,k_2}^{(0)}(\gamma_1\gamma_2)} \notag
    \\
    =E_{n_1,n_2}&(\gamma_1\gamma_2)_{min} \notag
    \\
    +(&A_1\sqrt{(n_1-1)n_1}+A_3(n_1-2)\sqrt{(n_1-1)n_1}+2Cn_2\sqrt{(n_1-1)n_1} \notag
    \\
    &+C\sqrt{(n_1-1)n_1})^2/(E_{n_1,n_2}^{(0)}(\gamma_1\gamma_2)-E_{n_1-2,n_2}^{(0)}(\gamma_1\gamma_2)) \notag
    \\ \notag
    \\
    +(&A_1\sqrt{(n_1+2)(n_1+1)}+A_3n_1\sqrt{(n_1+1)(n_1+2)}+2C\sqrt{(n_1+1)(n_1+2)}n_2 \notag
    \\
    &+C\sqrt{(n_1+1)(n_1+2)})^2/(E_{n_1,n_2}^{(0)}(\gamma_1\gamma_2)-E_{n_1+2,n_2}^{(0)}(\gamma_1\gamma_2)) \notag
    \\
    +&\frac{A_2^2}{E_{n_1,n_2}^{(0)}(\gamma_1\gamma_2)-E_{n_1-4,n_2}^{(0)}(\gamma_1\gamma_2)}(n_1-3)(n_1-2)(n_1-1)n_1 \notag
    \\
    +&\frac{A_2^2}{E_{n_1,n_2}^{(0)}(\gamma_1\gamma_2)-E_{n_1+4,n_2}^{(0)}(\gamma_1\gamma_2)}(n_1+4)(n_1+3)(n_1+2)(n_1+1) \notag
    \\ \notag
    \\ 
    +(&B_1\sqrt{(n_2-1)n_2}+B_3(n_2-2) \sqrt{(n_2-1)n_2}+2Cn_1\sqrt{(n_2-1)n_2} \notag
    \\
    &+C\sqrt{(n_2-1)n_2})^2/(E_{n_1,n_2}^{(0)}(\gamma_1\gamma_2)-E_{n_1,n_2-2}^{(0)}(\gamma_1\gamma_2)) \notag
    \\ \notag
    \\ 
    +(&B_1\sqrt{(n_2+2)(n_2+1)}+B_3n_2\sqrt{(n_2+2)(n_2+1)}+2Cn_1\sqrt{(n_2+2)(n_2+1)} \notag
    \\
    &+C\sqrt{(n_2+2)(n_2+1)})^2/(E_{n_1,n_2}^{(0)}(\gamma_1\gamma_2)-E_{n_1,n_2+2}^{(0)}(\gamma_1\gamma_2)) \notag
    \\
    +&\frac{B_2^2}{E_{n_1,n_2}^{(0)}(\gamma_1\gamma_2)-E_{n_1,n_2-4}^{(0)}(\gamma_1\gamma_2)}(n_2-3)(n_2-2)(n_2-1)n_2\notag
    \\
    +&\frac{B_2^2}{E_{n_1,n_2}^{(0)}(\gamma_1\gamma_2)-E_{n_1,n_2+4}^{(0)}(\gamma_1\gamma_2)}(n_2+4)(n_2+3)(n_2+2)(n_2+1) \notag
    \end{align}
    
    \begin{align}
    &+\frac{C^2}{E_{n_1,n_2}^{(0)}(\gamma_1\gamma_2)-E_{n_1-2,n_2-2}^{(0)}(\gamma_1\gamma_2)}(n_1-1)n_1(n_2-1)n_2 \notag
    \\
    &+\frac{C^2}{E_{n_1,n_2}^{(0)}(\gamma_1\gamma_2)-E_{n_1-2,n_2+2}^{(0)}(\gamma_1\gamma_2)}(n_1-1)n_1(n_2+2)(n_2+1) \notag
    \\
    &+\frac{C^2}{E_{n_1,n_2}^{(0)}(\gamma_1\gamma_2)-E_{n_1+2,n_2-2}^{(0)}(\gamma_1\gamma_2)}(n_1+2)(n_1+1)(n_2-1)n_2 \notag
    \\
    &+\frac{C^2}{E_{n_1,n_2}^{(0)}(\gamma_1\gamma_2)-E_{n_1+2,n_2+2}^{(0)}(\gamma_1\gamma_2)}(n_1+2)(n_1+1)(n_2+2)(n_2+1) \notag
    \end{align}

	\subsection{Degenerate state}
	It is obvious that there exists double degeneracy in the strongly coupled double harmonic oscillator system if the quasiparticle number $n_1$ is equal to $k_2$, $n_2$ is equal to $k_1$, and $n_1$ is not equal to $n_2$. Checking the elements in the pertubation V, we find that only $A_{\gamma_1}^{\dagger2}A_{\gamma_2}^{2}$ and $A_{\gamma_1}^{2}A_{\gamma_2}^{\dagger2}$ can make the numerator not zero when the denominator is zero in the second-order energy correction formula. Therefore, the modified formula of nondegenerate perturbation energy is not applicable. Next, we will use degenerate perturbation theory to deal with this case.\\
	
	$H^{\prime}=A_{\gamma_1}^{\dagger2}A_{\gamma_2}^{2}+A_{\gamma_1}^{2}A_{\gamma_2}^{\dagger2}$ has two states, $\psi_1=|n_1n_2\rangle$ and $\psi_2=|n_1+2,n_2-2\rangle$.\\
	
	Calculate
	\begin{align}
		H_{11}^{\prime}&=\langle \psi_1|H^{\prime}|\psi_1\rangle \\
		&=0 \notag
	\end{align}
	
	\begin{align}
	    H_{12}^{\prime}&=\langle \psi_1|H^{\prime}|\psi_2\rangle
	    \\
	    &=C\sqrt{(n_1+2)(n_1+1)(n_2-1)n_2} \notag
	\end{align}
	
	\begin{align}
     	H_{21}^{\prime}&=\langle \psi_2|H^{\prime}|\psi_1\rangle
     	\\
     	&=C\sqrt{(n_1+2)(n_1+1)(n_2-1)n_2} \notag
	\end{align}
	
	\begin{align}
	    H_{22}^{\prime}&=\langle \psi_2|H^{\prime}|\psi_2\rangle
	    \\
	    &=0 \notag
	\end{align}

\begin{center}
	$H^{\prime}$=  $\begin{pmatrix}
					H_{11}^{\prime} & H_{12}^{\prime} \\
					H_{21}^{\prime} & H_{22}^{\prime}
				\end{pmatrix}$
\end{center}

	\begin{align}
	det(H^{\prime}-EI)=0
	\end{align}
	
	\begin{align}
	\Rightarrow E_{n_1n_2}^{(1)}=C\sqrt{(n_1+2)(n_1+1)n_2(n_2-1)} 
	\end{align}
	
	So we have the energy correction formula for the double degenerate state.
	
	\begin{align}
		E_{n_1,n_2}(\gamma_1\gamma_2)_{2}=E_{n_1,n_2}(\gamma_1\gamma_2)_{min}+E_{n_1n_2}^{(1)}
	\end{align}
	
	\subsection{The reason of convergence}
	
	The eigenenergy calculated by the time-independent perturbation is
	
	\begin{align}
	E_n=E_n^{(0)}+\langle n^{(0)}|V|n^{(0)}\rangle+\sum_{k \neq n}\dfrac{|\langle k^{(0)}|V|n^{(0)}\rangle|^2}{E_n^{(0)}-E_k^{(0)}}+...
	\end{align}
	
	\noindent where $E_n^{(0)}$ is the $n^{th}$ unperturbed eigenenergy defined by the $E_n{(\gamma)}_{min}$, $|n^{(0)}\rangle$ is the $n^{th}$ unperturbed eigenstate. We will use the time-independent perturbation to explain why the adaptive perturbation theory converges. \\
	
	It can be seen from equation 7 and equation 14 that when Hamiltonian \textbf{\textit{$H_{0}$}} acts on $|n_{\gamma_1},n_{\gamma_2}\rangle$ state, when the quasiparticle number \textbf{\textit{n}} is large enough, the dominant factor is $\frac{\lambda}{4\gamma^2}n(n-1)$, which contributes the square of \textbf{\textit{n}}, so $E_n^{(0)}-E_k^{(0)}$ provides $n^2$. When the gap between \textbf{\textit{$n_{1}$}} and \textbf{\textit{$n_{2}$}} is small and their values are large enough, the coupling term   $\frac{\lambda}{4\gamma_1\gamma_2}(2n_1+1)(2n_2+1)$ also contributes \textbf{\textit{$n^2$}}. While the gap between \textbf{\textit{$n_{1}$}} and \textbf{\textit{$n_{2}$}} and one value of them are large enough, it only contributes \textbf{\textit{n}}. The $ n^2 $ provided by $E_n^{(0)}-E_k^{(0)}$ appears in the denominator. For more advanced energy correction, more multiplications of the $E_n^{(0)}-E_k^{(0)}$ will be found. The calculation of the higher-order
	term will be suppressed by the multiplications of the $E_n^{(0)}-E_k^{(0)}$. So the adaptive perturbation theory will converge. In additon,  I. G. Halliday and P. Suranyi give a proof that the fact that the unperturbed energies are quadratic in $A^{\dagger}_{\gamma}A_{\gamma}$ leads to a convergent perturbation expansion[5,6].

	\section{Results}
	
	\subsection{Numerical comparison of Leading-Order}
	The following five tables record the deviations of the \textbf{$\textit{E}_{n_1,n_2}(\gamma_1\gamma_2)_{min}$} and the numerical solution when \textbf{\textit{$\lambda$}} = 0.5, 1, 2, 8, 16.
	
	\begin{center}
		\begin{tabular}{|c|c|c|c|c|}
			\hline
			$\textit{n}_{1}$ & $\textit{n}_{2}$ &$\textit{E}_{n_1,n_2}(\gamma_1\gamma_2)_{min}$ &Numerical Solution &Deviation\\
			\hline
			0 &0 &0.90806 &0.89876 &1.0348\% \\
			\hline
			0 &1 &1.98970 &1.97417 &0.7867\% \\
			\hline
			0 &2 &3.20303 &3.32503 &3.6691\% \\
			\hline
			0 &3 &4.52592 &4.39998 &2.8623\% \\
			\hline
			1 &1 &3.34985 &3.04935 &9.8546\% \\
			\hline
			1 &2 &4.77246 &4.85193 &1.6379\% \\
			\hline
			1 &3 &6.26398 &6.51984 &3.9243\% \\
			\hline
			2 &2 &6.37404 &5.75024 &10.8482\% \\
			\hline
		\end{tabular}
	\end{center}
	
	\begin{center}
		{\tiny \textbf{Table 1: The comparison between the $\textit{E}_{n_1,n_2}(\gamma_1\gamma_2)_{min}$ and the numerical solutions for the \textit{$\lambda$} = 0.5}} 
	\end{center}
	\hspace*{\fill} 
	
	\begin{center}
		\begin{tabular}{|c|c|c|c|c|}
			\hline
			$\textit{n}_{1}$ & $\textit{n}_{2}$ &$\textit{E}_{n_1,n_2}(\gamma_1\gamma_2)_{min}$ &Numerical Solution &Deviation\\
			\hline
			0 &0 &1.19260 &1.18211 &0.8874\% \\
			\hline
			0 &1 &2.59473 &2.57669 &0.7001\% \\
			\hline
			0 &2 &4.15752 &4.30631 &3.4552\% \\
			\hline
			0 &3 &5.85480 &5.70014 &2.7133\% \\
			\hline
			1 &1 &4.33579 &3.97089 &9.1894\% \\
			\hline
			1 &2 &6.15608 &6.25446 &1.5730\% \\
			\hline
			1 &3 &8.06224 &8.37789 &3.7677\% \\
			\hline
			2 &2 &8.19710 &7.42878 &10.3425\% \\
			\hline
		\end{tabular}
	\end{center}
	
	\begin{center}
		{\tiny \textbf{Table 2: The comparison between the $\textit{E}_{n_1,n_2}(\gamma_1\gamma_2)_{min}$ and the numerical solutions for the \textit{$\lambda$} = 1}} 
	\end{center}
	\hspace*{\fill} 
	
	\begin{center}
		\begin{tabular}{|c|c|c|c|c|}
			\hline
			$\textit{n}_{1}$ & $\textit{n}_{2}$ &$\textit{E}_{n_1,n_2}(\gamma_1\gamma_2)_{min}$ &Numerical Solution &Deviation\\
			\hline
			0 &0 &1.57746 &1.56590 &0.7382\% \\
			\hline
			0 &1 &3.40543 &3.38480 &0.6095\% \\
			\hline
			0 &2 &5.42793 &5.60795 &3.2101\% \\
			\hline
			0 &3 &7.61451 &7.42564 &2.5435\% \\
			\hline
			1 &1 &5.64259 &5.20310 &8.4467\% \\
			\hline
			1 &2 &7.98009 &8.10113 &1.4941\% \\
			\hline
			1 &3 &10.42420 &10.81140 &3.5814\% \\
			\hline
			2 &2 &10.58840 &9.64721 &9.7561\% \\
			\hline
		\end{tabular}
	\end{center}
	
	\begin{center}
		{\tiny \textbf{Table 3: The comparison between the $\textit{E}_{n_1,n_2}(\gamma_1\gamma_2)_{min}$ and the numerical solutions for the \textit{$\lambda$} = 2}} 
	\end{center}
	\hspace*{\fill} 
	
	\begin{center}
		\begin{tabular}{|c|c|c|c|c|}
			\hline
			$\textit{n}_{1}$ & $\textit{n}_{2}$ &$\textit{E}_{n_1,n_2}(\gamma_1\gamma_2)_{min}$ &Numerical Solution &Deviation\\
			\hline
			0 &0 &2.82393 &2.81087 &0.4646\% \\
			\hline
			0 &1 &5.99335 &5.96769 &0.4300\% \\
			\hline
			0 &2 &9.44007 &9.69528 &2.6323\% \\
			\hline
			0 &3 &13.1251 &12.84880 &2.1504\% \\
			\hline
			1 &1 &9.74173 &9.12281 &6.7843\% \\
			\hline
			1 &2 &13.64890 &13.82700 &1.2881\% \\
			\hline
			1 &3 &17.71840 &18.28620 &3.1051\% \\
			\hline
			2 &2 &17.95570 &16.57220 &8.3483\% \\
			\hline
		\end{tabular}
	\end{center}
	
	\begin{center}
		{\tiny \textbf{Table 4: The comparison between the $\textit{E}_{n_1,n_2}(\gamma_1\gamma_2)_{min}$ and the numerical solutions for the \textit{$\lambda$} = 8}} 
	\end{center}
	\hspace*{\fill} 
	
	\begin{center}
		\begin{tabular}{|c|c|c|c|c|}
			\hline
			$\textit{n}_{1}$ & $\textit{n}_{2}$ &$\textit{E}_{n_1,n_2}(\gamma_1\gamma_2)_{min}$ &Numerical Solution &Deviation\\
			\hline
			0 &0 &3.82278 &3.80935 &0.3526\% \\
			\hline
			0 &1 &8.04283 &8.01471 &0.3509\% \\
			\hline
			0 &2 &12.58820 &12.88570 &2.3088\% \\
			\hline
			0 &3 &17.41650 &17.08550 &1.9373\% \\
			\hline
			1 &1 &12.93920 &12.21720 &5.9097\% \\
			\hline
			1 &2 &18.03310 &18.24440 &1.1582\% \\
			\hline
			1 &3 &23.32510 &24.00070 &2.8149\% \\
			\hline
			2 &2 &23.60610 &21.94960 &7.5468\% \\
			\hline
		\end{tabular}
	\end{center}
	
	\begin{center}
		{\tiny \textbf{Table 5: The comparison between the $\textit{E}_{n_1,n_2}(\gamma_1\gamma_2)_{min}$ and the numerical solutions for the \textit{$\lambda$} = 16}} 
	\end{center}
	\hspace*{\fill} 
	
	\subsection{Numerical comparison of 2nd-Order}
	The following tables record the deviations between the $E_{n_1,n_2}(\gamma_1\gamma_2)_2$ and the numerical solution.
	
	\begin{center}
		\begin{tabular}{|c|c|c|c|c|}
			\hline
			$\textit{n}_{1}$ & $\textit{n}_{2}$ &$\textit{E}_{n_1,n_2}(\gamma_1\gamma_2)_{2}$ &Numerical Solution &Deviation\\
			\hline
			0 &0 &0.899293 &0.89876 &0.0593\% \\
			\hline
			0 &1 &1.97477 &1.97417 &0.0302\% \\
			\hline
			0 &2 &3.32385 &3.32503 &0.0356\% \\
			\hline
			0 &3 &4.43047 &4.39998 &0.6929\% \\
			\hline
			1 &1 &3.32569 &3.04935 &9.0623\% \\
			\hline
			1 &2 &4.94093 &4.85193 &1.8344\% \\
			\hline
			1 &3 &6.54223 &6.51984 &0.3434\% \\
			\hline
			2 &2 &6.13216 &5.75024 &6.6419\% \\
			\hline
		\end{tabular}
	\end{center}
	
	\begin{center}
		{\tiny \textbf{Table 6: The deviation between the $E_{n_1,n_2}(\gamma_1\gamma_2)_2$ and the numerical solutions for the \textit{$\lambda$} = 0.5.}} 
	\end{center}

	\hspace*{\fill} 
	
	\begin{center}
		\begin{tabular}{|c|c|c|c|c|}
			\hline
			$\textit{n}_{1}$ & $\textit{n}_{2}$ &$\textit{E}_{n_1,n_2}(\gamma_1\gamma_2)_{2}$ &Numerical Solution &Deviation\\
			\hline
			0 &0 &1.18266 &1.18211 &0.0465\% \\
			\hline
			0 &1 &2.57743 &2.57669 &0.0287\% \\
			\hline
			0 &2 &4.30551 &4.30631 &0.0185\% \\
			\hline
			0 &3 &5.73462 &5.70014 &0.6048\% \\
			\hline
			1 &1 &4.30734 &3.97089 &8.4730\% \\
			\hline
			1 &2 &6.36156 &6.25446 &1.7123\% \\
			\hline
			1 &3 &8.40525 &8.37789 &0.3266\% \\
			\hline
			2 &2 &8.4964 &7.42878 &14.3713\% \\
			\hline
		\end{tabular}
	\end{center}
	
	\begin{center}
		{\tiny \textbf{Table 7: The deviation between the $E_{n_1,n_2}(\gamma_1\gamma_2)_2$ and the numerical solutions for the \textit{$\lambda$} = 1.}} 
	\end{center}

	\hspace*{\fill} 
	
	\begin{center}
		\begin{tabular}{|c|c|c|c|c|}
			\hline
			$\textit{n}_{1}$ & $\textit{n}_{2}$ &$\textit{E}_{n_1,n_2}(\gamma_1\gamma_2)_{2}$ &Numerical Solution &Deviation\\
			\hline
			0 &0 &1.56646 &1.56590 &0.0359\% \\
			\hline
			0 &1 &3.38578 &3.38480 &0.0289\% \\
			\hline
			0 &2 &5.60798 &5.60795 &$4.68072*10^{-4}$\% \\
			\hline
			0 &3 &7.46358 &7.42564 &0.5109\% \\
			\hline
			1 &1 &5.60964 &5.20310 &7.8134\% \\
			\hline
			1 &2 &8.22898 &8.10113 &1.5782\% \\
			\hline
			1 &3 &10.8448 &10.81140 &0.3089\% \\
			\hline
			2 &2 &10.9571 &9.64721 &13.5777\% \\
			\hline
		\end{tabular}
	\end{center}
	
	\begin{center}
		{\tiny \textbf{Table 8: The deviation between the $E_{n_1,n_2}(\gamma_1\gamma_2)_2$ and the numerical solutions for the \textit{$\lambda$} = 2.}} 
	\end{center}

	\hspace*{\fill} 
	
	\begin{center}
		\begin{tabular}{|c|c|c|c|c|}
			\hline
			$\textit{n}_{1}$ & $\textit{n}_{2}$ &$\textit{E}_{n_1,n_2}(\gamma_1\gamma_2)_{2}$ &Numerical Solution &Deviation\\
			\hline
			0 &0 &2.81164 &2.81087 &0.0272\% \\
			\hline
			0 &1 &5.96984 &5.96769 &0.0361\% \\
			\hline
			0 &2 &9.69959 &9.69528 &0.0444\% \\
			\hline
			0 &3 &12.8903 &12.84880 &0.3231\% \\
			\hline
			1 &1 &9.70024 &9.12281 &6.3295\% \\
			\hline
			1 &2 &14.0050 &13.82700 &1.2877\% \\
			\hline
			1 &3 &18.3376 &18.28620 &0.2813\% \\
			\hline
			2 &2 &18.5051 &16.57220 &11.6635\% \\
			\hline
		\end{tabular}
	\end{center}
	
	\begin{center}
		{\tiny \textbf{Table 9: The deviation between the $E_{n_1,n_2}(\gamma_1\gamma_2)_2$ and the numerical solutions for the \textit{$\lambda$} =8.}} 
	\end{center}
	
	\hspace*{\fill} 
	
	\begin{center}
		\begin{tabular}{|c|c|c|c|c|}
			\hline
			$\textit{n}_{1}$ & $\textit{n}_{2}$ &$\textit{E}_{n_1,n_2}(\gamma_1\gamma_2)_{2}$ &Numerical Solution &Deviation\\
			\hline
			0 &0 &3.81047 &3.80935 &0.0293\% \\
			\hline
			0 &1 &8.01829 &8.01471 &0.0447\% \\
			\hline
			0 &2 &12.8946 &12.88570 &0.0692\% \\
			\hline
			0 &3 &17.1268 &17.08550 &0.2418\% \\
			\hline
			1 &1 &12.8945 &12.21720 &5.5436\% \\
			\hline
			1 &2 &18.4528 &18.24440 &1.1424\% \\
			\hline
			1 &3 &24.0664 &24.00070 &0.2793\% \\
			\hline
			2 &2 &24.2689 &21.94960 &10.5665\% \\
			\hline
		\end{tabular}
	\end{center}
	
	\begin{center}
		{\tiny \textbf{Table 10: The deviation between the $E_{n_1,n_2}(\gamma_1\gamma_2)_2$ and the numerical solutions for the \textit{$\lambda$} = 16.}} 
	\end{center}

	\begin{figure}[p]
		\centering
		\includegraphics[scale=0.6]{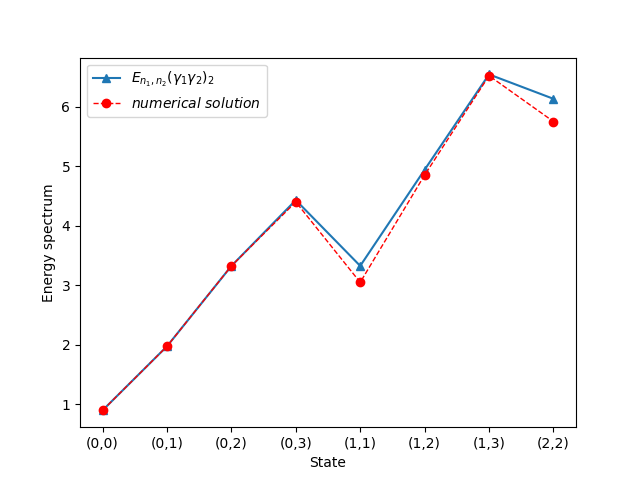}
		\caption{The state-energy spectrum diagram of the comparison between $E_{n_1,n_2}(\gamma_1\gamma_2)_2$ and the numerical solution when the coupling constant \textit{$\lambda$} is 0.5.}
	\end{figure}

	\begin{figure}[p]
		\centering
		\includegraphics[scale=0.6]{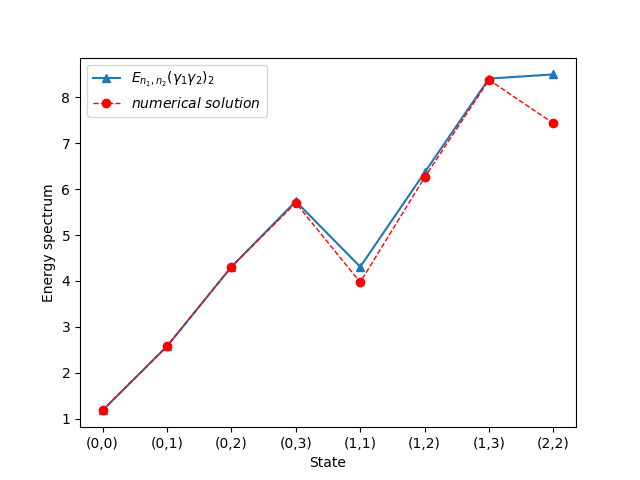}
		\caption{The state-energy spectrum diagram of the comparison between $E_{n_1,n_2}(\gamma_1\gamma_2)_2$ and the numerical solution when the coupling constant \textit{$\lambda$} is 1.}
	\end{figure}
	
	\begin{figure}[p]
		\centering
		\includegraphics[scale=0.6]{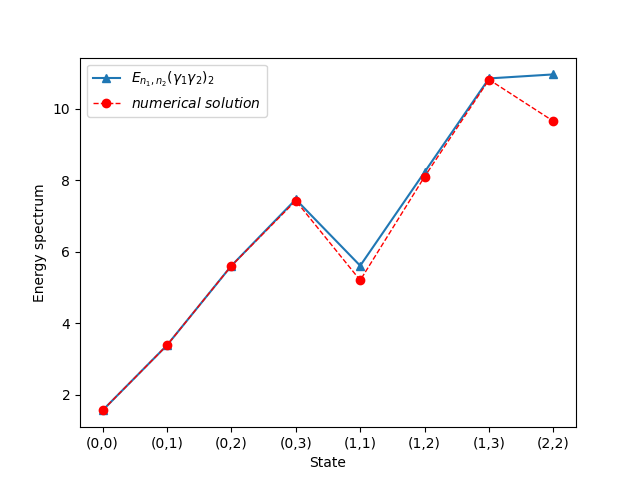}
		\caption{The state-energy spectrum diagram of the comparison between $E_{n_1,n_2}(\gamma_1\gamma_2)_2$ and the numerical solution when the coupling constant \textit{$\lambda$} is 2.}
	\end{figure}

	\begin{figure}[p]
		\centering
		\includegraphics[scale=0.6]{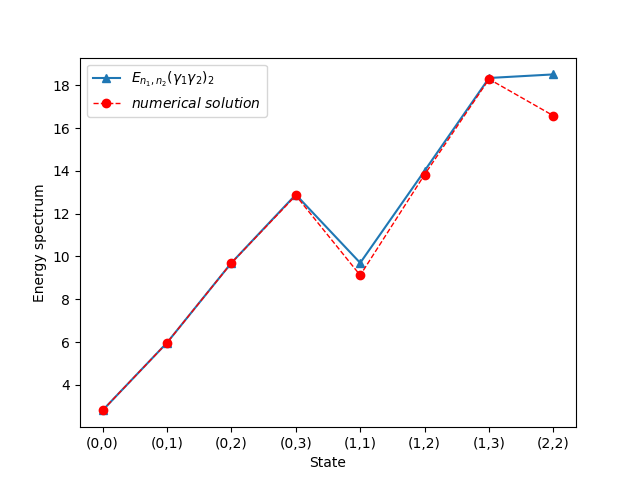}
		\caption{The state-energy spectrum diagram of the comparison between $E_{n_1,n_2}(\gamma_1\gamma_2)_2$ and the numerical solution when the coupling constant \textit{$\lambda$} is 8.}
	\end{figure}

	\begin{figure}[h]
		\centering
		\includegraphics[scale=0.6]{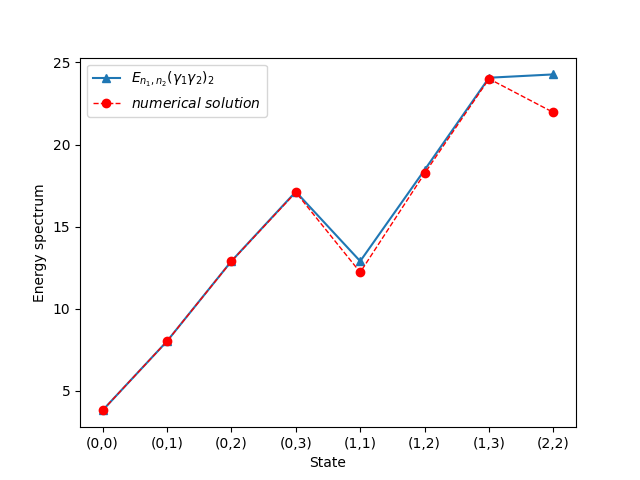}
		\caption{The state-energy spectrum diagram of the comparison between $E_{n_1,n_2}(\gamma_1\gamma_2)_2$ and the numerical solution when the coupling constant \textit{$\lambda$} is 16.}
	\end{figure}

	\subsection{The analysis and explanation}	
	 All the tables above record the difference between the corresponding leading-order energy or the energy after second-order correction and the corresponding real energy numerical solution when the strongly coupled double harmonic oscillator system is in the state represented by $|(n_1,n_2)\rangle$. \\
	 
	 The deviations between the leading-order and the numerical solution show that as the gap between the quasiparticle number \textbf{\textit{$n_{1}$}} and \textbf{\textit{$n_{2}$}} decreases, which means the excitation level of the system is lower; or the coupling constant $\lambda$ increases, which means the interaction between the two harmonic oscillators is enhanced, the gap between the exact solutions in the strongly coupling region of double harmonic oscillator system and the numerical solutions becomes smaller. Overall, the numerical gap is about 1\% to 3\%, which is not a bad result. When \textbf{\textit{$n_{1}$}} and \textbf{\textit{$n_{2}$}} are equal: if they are not zero, that is, the strong coupling double harmonic oscillator system is not in the ground state, the gap between the leading-order and the numerical solution will be quite large; only when the strong coupling double harmonic oscillator system is in the ground state, the gap between the leading-order and the numerical solution is small, the exact solution of the double harmonic oscillator quantum system in the strong coupling field calculated by the adaptive perturbation theory is close to the real value of energy. \\
	 
	 The tables recording the deviations between the second-order and the numerical solution show that when the quasiparticle number \textbf{\textit{$n_{1}$}} in the strongly coupled double harmonic oscillator system is not equal to the quasiparticle number \textbf{\textit{$n_{2}$}}, the gap between the two is quite small. The analytic solution of the double harmonic oscillator quantum system in the strong coupling field obtained by the adaptive perturbation theory is quite close to the real value of energy. In most cases, the correlation error is less than 1\%. Generally speaking, the gap between the quasiparticle number \textbf{\textit{$n_{1}$}}  and \textbf{\textit{$n_{2}$}} is positively correlated with the gap between the analytical solution and the relevant numerical solution of the double harmonic oscillator quantum system in the strong coupling field. That is to say, the higher the excitation degree of the system, the greater the error of the adaptive perturbation theory. The coupling constant $\lambda$ between the two harmonic oscillators is negatively correlated with the gap between the analytical solution and the numerical solution. In other words, the stronger the interaction between the two harmonic oscillators, the smaller the error of the adaptive perturbation theory. \\
	 
	 To sum up, the adaptive perturbation theory is effective for the double harmonic oscillator quantum system in the field of strong coupling. The adaptive perturbation theory can be applied to the latter. The adaptive perturbation theory breaks through the limitation of the traditional perturbation theory.

	\section{Outlook}
	In the real world, we can't never encounter the strong coupling problem, so it is necessary to develop the theory suitable for the strong coupling field. Because boson quantum mechanics is interlinked with quantum field theory in perturbation problems and theoretical formulas, the example in this paper can be easily extended to boson quantum field theory. Now, since the adaptive perturbation theory can be applied to the strongly coupled double harmonic oscillator quantum system, we have the reason to think that the prospect of extending the adaptive perturbation theory to the strongly coupled quantum field theory is bright, which can provide more clues for other open problems: for example, the lattice method is often used to study the problem of strong coupling quantum field theory, but due to the limitation of continuum, it is difficult to determine the correctness of the lattice method. Now we can use the adaptive perturbation theory to verify the correctness of the former answer: we can use adaptive perturbation theory to study quarks and gluons in quantum chromodynamics, study phase transition in the field of strong coupling, and explore the critical point in quantum chromodynamics. In addition, we can extend the adaptive perturbation theory to the scalar field theory to study the Higgs particle mechanism and the source of mass.
	
	\section{References}
	[1] C. T. Ma, “Accurate Study from
	Adaptive Perturbation Method,” \\
	(https://arxiv.org/abs/2007.09080v3)
	
	\hspace*{\fill}
	
	[2] M. Weinstein, “Adaptive perturbation theory. I. Quantum mechanics,” \\
	(https://arxiv.org/abs/hep-th/0510159)
	
	\hspace*{\fill}
	
	[3] C. T. Ma, “Second-Order Perturbation in
	Adaptive Perturbation Method,” \\
	(https://arxiv.org/abs/2004.00842v3)
	
	\hspace*{\fill}
	
	[4] C. T. Ma, “Adaptive Perturbation Method in Bosonic Quantum Mechanics,”
	(https://arxiv.org/abs/1911.08211)
	
    \hspace*{\fill} 
	
	[5] I. G. Halliday and P. Suranyi, Phys. Lett. B 85, 421 (1979) \\
	
	\hspace*{\fill}
	
	[6] I. G. Halliday and P. Suranyi, Phys. Rev. D 21, 1529 (1980)
\end{document}